\documentclass[doublecol]{epl2} 
\usepackage{amssymb}
\usepackage{subfigure}
\usepackage{graphicx}% Include figure files
\usepackage{dcolumn,color}% Align table columns on decimal point
\usepackage{bm}
\usepackage{caption}
\usepackage{epstopdf}
\usepackage{subeqnarray}
\usepackage{epsfig}
\usepackage{color,framed}
\usepackage{amsmath}
\usepackage[table]{xcolor}
\usepackage{cases}
\usepackage{mathrsfs}

\title{Scalar perturbation of gravitating double-kink solutions}
\shorttitle{Scalar perturbation of gravitating double-kink solutions} %Insert here a short version of the title if it exceeds 70 characters

\author{Jun Feng and Yuan Zhong\footnote{E-mail: zhongy@mail.xjtu.edu.cn (corresponding author)} }
\shortauthor{J. Feng and Y. Zhong }

\institute{                    
School of Physics, Xi'an Jiaotong University, Xi'an 710049, People's Republic of China
}
\abstract{
In this letter, a two-dimensional (2D) gravity-scalar model is studied. This model supports interesting double-kink solutions, and the corresponding metric solutions can be derived analytically. Depending on a tunable parameter $c$, the metric can be symmetric or asymmetric. The Schr\"odinger-like equation for normal modes of the physical linear perturbation is derived. As $c$ varies, the effective potential can have one or two singular barriers. If $c$ is larger than a critical value, the zero mode will be normalizable, despite of the appearance of a strong repulsive singularity. The double-kink solution is always stable against linear perturbations.}

\begin{document}

\maketitle

%---------------------
\section{Introduction}
%---------------------
Kink is a simple type of topological soliton solution, which connects two different vacua of a scalar field model. Two representative kink models are the sine-Gordon model and the $\phi^4$ model in 2D Minkowski space-time~\cite{Rajaraman1975}. The applications of kink solutions range from condensed matter physics~\cite{BishopSchneider1978} to high energy physics~\cite{Vachaspati2006}. Especially, self-gravitating kink solutions of some 5D models can be regarded as thick brane worlds, on which both matter and gravity fields can be trapped dynamically~\cite{DeWolfeFreedmanGubserKarch2000,Gremm2000,CsakiErlichHollowoodShirman2000,KoleyKar2005}, see Refs.~\cite{DzhunushalievFolomeevMinamitsuji2010,Liu2018} for comprehensive reviews.

Although most thick brane models are based on single kink solutions, there are also thick brane models based on gravitating double-kink solutions, either in Einstein gravity~\cite{Campos2002,BazeiaFurtadoGomes2004} or in modified gravity theories~\cite{ZhongZhongZhangLiu2018,GuoZhongYangSuiEtAl2020,XiangZhongXieZhao2020,ChenGuoLiu2021,XieFuSuiZhaoEtAl2021,XuChenZhangLiu2022}. Gravitating double-kink solutions are different from the single kink ones in many aspects. For example, the energy density of a double-kink usually has more peaks than a single kink. Besides, there might be some resonant modes in the tensor perturbation spectrum of a 5D gravitating double-kink~\cite{GuoZhongYangSuiEtAl2020,XiangZhongXieZhao2020,ChenGuoLiu2021,XieFuSuiZhaoEtAl2021,XuChenZhangLiu2022}.

It is natural to ask, how a gravitating double-kink would be different from a single kink when the scalar perturbations are taken into consideration. Unfortunately, this issue was seldom discussed. Partly because the scalar perturbation issue of a 5D thick brane models is, in general, much more complicated than the tensor one. Recently, it was found that some 2D gravity-scalar models can reflect many properties of 5D thick brane models, including the scalar perturbation spectrum~\cite{Zhong2021,ZhongLiLiu2021}. The linear perturbation issue of a 2D model is much easier than the 5D case. Not just because we only have scalar perturbations in two dimensions, in fact, the linear perturbation equations can be easily derived with the help of Mathematica programs~\cite{Zhong2021}.

For simplicity, in the present work we focus on the scalar perturbation issue of a class of 2D gravitating double-kink solutions. We will first construct a 2D gravity-scalar field model, which support double-kink solutions, and then discuss the stability issue following the procedures of Refs.~\cite{Zhong2021,ZhongLiLiu2021,Zhong2021b}.

%---------------------
\section{The model and double-kink solutions}
In two dimensions, the Einstein-Hilbert action is topologically invariant, and cannot describe the dynamics of gravity. Therefore, one has to consider some modified theories of gravity, such as scalar-tensor gravity~\cite{Jackiw1985,Henneaux1985,MannMorsinkSikkemaSteele1991,CallanGiddingsHarveyStrominger1992,Stoetzel1995}, $f(R)$ gravity~\cite{Schmidt1999,NojiriOdintsov2020}, and so on~\cite{Brown1988,NojiriOdintsov2001d,GrumillerKummerVassilevich2002}. Among these 2D gravity theories, the following one has relatively simple field equations~\cite{MannMorsinkSikkemaSteele1991,Stoetzel1995}:
\begin{equation}\label{1}
S=\frac{1}{\kappa} \int d^{2} x \sqrt{-g}\left[-\frac{1}{2} \nabla^{\mu} \varphi \nabla_{\mu} \varphi+\varphi R+\kappa \mathcal{L}_m\right],
\end{equation}
where $\kappa$ is the gravitational coupling constant, $\varphi$ is a dilaton field,  and 
\begin{equation}
\mathcal{L}_m=-\frac12g^{\mu\nu}\nabla_\mu\phi\nabla_\nu\phi-V(\phi),
\end{equation}
is the Lagrangian density of a real scalar matter field.

After variations, one would obtain three dynamical equations, namely, the Einstein equation
\begin{eqnarray}
\label{STeq1}
&&\nabla_{\mu} \varphi \nabla_{\nu} \varphi-\frac{1}{2} g_{\mu \nu}\left(\nabla^{\lambda} \varphi \nabla_{\lambda} \varphi+4 \nabla_{\lambda} \nabla^{\lambda} \varphi\right)\nonumber\\
&+&2 \nabla_{\mu} \nabla_{\nu} \varphi=-\kappa T_{\mu \nu},
\end{eqnarray}
the dilaton equation
\begin{equation}
\label{STeq2}
\nabla^{\lambda} \nabla_{\lambda} \varphi+R=0,
\end{equation}
and the scalar equation
\begin{eqnarray}
\label{STeq3}
 \nabla_{\lambda} \nabla^{\lambda} \phi = \frac{\partial V}{ \partial{\phi}}.
\end{eqnarray}
The energy-momentum tensor in Eq.~\eqref{STeq1} is defined by
\begin{equation}
T_{\mu \nu}=g_{\mu \nu}\mathcal{L}_m+  \nabla _{\mu}\phi \nabla _{\nu}\phi.
\end{equation}

As in Refs.~\cite{Stoetzel1995,Zhong2021,ZhongLiLiu2021}, we look for static solutions with the following metric:
\begin{equation}
\label{metricXCord}
  ds^2=-e^{2A(x)}dt^2+dx^2,
\end{equation}
where $A(x)$ is the warp factor. With this metric, the dilaton equation \eqref{STeq2} gives an  algebraic relation between the dilaton field and the warp factor~\cite{Stoetzel1995,Zhong2021,ZhongLiLiu2021}
\begin{eqnarray}
\varphi=2A.
\end{eqnarray}
Using this relation, the Einstein equations can be simplified in the following form:
\begin{eqnarray}
\label{eqEin1}
-4 \partial_x^2 A&=&\kappa   (\partial_x\phi)^2, \\
\label{eqEin2}
2 \partial_x^2 A +2 \left(\partial_x A\right)^2&=&-\kappa V,
\end{eqnarray}
from which the scalar field equation
\begin{equation}
\partial_x A \partial_x \phi+\partial_x^2 \phi
=V_{\phi }
\end{equation}
 can be derived. Therefore, to find a solution of the system, we only need to solve Eqs.~\eqref{eqEin1} and \eqref{eqEin2}.
\begin{figure*}
\begin{center}
\centerline{\includegraphics[width=16cm]{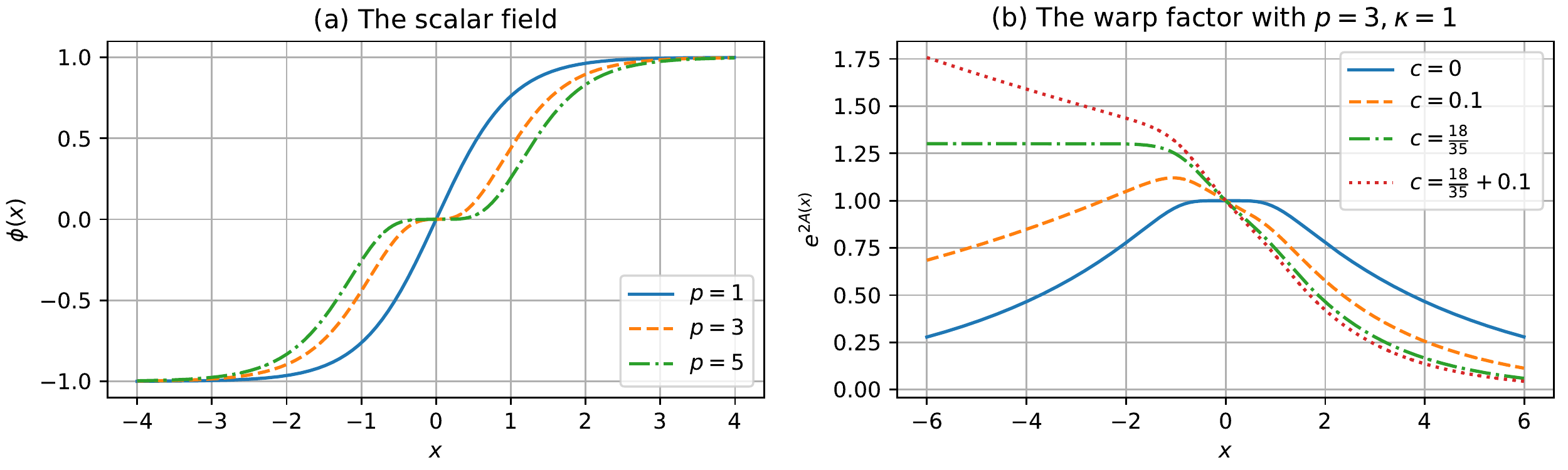}}
\caption{Plots of the scalar field $\phi(x)$ and the warp factor $e^{2A(x)}$.}
\label{Fig1}
\end{center}
\end{figure*}

As shown in Refs.~\cite{Zhong2021,ZhongLiLiu2021}, the Einstein equations~\eqref{eqEin1} and \eqref{eqEin2} have a simple first-order formalism, from which exact solutions can be constructed. To see this, we first introduce the superpotential function $W(\phi)$ by assuming
\begin{equation}
\label{EqFirsForm1}
\partial_x\phi=\frac{d W}{d \phi}.
\end{equation}
Substituting this equation into Eqs.~\eqref{eqEin1} and \eqref{eqEin2} we immediately obtain
\begin{equation}
\label{EqFirsForm2}
\partial_x A=-\frac14\kappa W,
\end{equation}
and
\begin{equation}
\label{EqFirsForm3}
V=\frac{1}{2} W_\phi^2-\frac{1}{8} \kappa  W^2.
\end{equation}
Given an appropriate superpotential function, the first-order equations \eqref{EqFirsForm1}-\eqref{EqFirsForm3} can be analytically solved.
Some examples of single kink solutions can be found in Ref.~\cite{Zhong2021}.

To obtain double-kink solutions, we adopt the superpotential proposed by Bazeia et. al.~\cite{BazeiaMenezesMenezes2003,BazeiaFurtadoGomes2004}:
\begin{equation}
W=
\frac{p^2}{2p-1} \phi ^{\frac{2p-1}p}-  \frac{p^2}{2p+1} \phi ^{\frac{2p+1}p}+c,
\end{equation}
where $p$ is a positive integer, and $c>0$ is a constant parameter. For this superpotential, the scalar field solution takes the following form:
\begin{eqnarray}
\label{solKink}
\phi(x)&=&\tanh^p (x).
\end{eqnarray}
When $p=1$, this solution reduces to the gravitating kink of Ref.~\cite{Zhong2021}, where the warp factor reads
\begin{eqnarray}
A(x)=-\frac{1}{24} \kappa  \left[6 c x+4 \ln (\cosh (x) )+\tanh ^2(x)\right].
\end{eqnarray}
If $p=3, 5, \cdots$, the solution in Eq.~\eqref{solKink} describes double-kink solutions~\cite{BazeiaMenezesMenezes2003,BazeiaFurtadoGomes2004}, see Fig.~\ref{Fig1}-(a).  For simplicity, we only discuss the case with $p=3$, and the corresponding warp factor solution reads
\begin{eqnarray}
\label{SolWarp}
A(x)&=&-\frac{1}{280} \kappa  \big[70 c x+36 \ln (\cosh (x))-18 \tanh ^2(x)\nonumber\\
&-&9 \tanh ^4(x)+15 \tanh ^6(x)\big].
\end{eqnarray}
The asymptotic behavior of the warp factor is
\begin{equation}
\label{eqAsymA}
\lim_{x\to\pm\infty}A=\frac{\kappa}{140}   [6 -  (18\pm 35c ) \left| x\right| +18 \ln (2)].
\end{equation}
Obviously, if $c=0$ the warp factor describes two symmetrically connected asymptotic AdS$_2$ spaces. If $c>0$ the warp factor is not symmetric any more. Moreover, there is a critical case, $c=c_{\textrm{cr}}=\frac{18}{35}\approx 0.5143$, for which the left side of the double-kink is an asymptotic Minkowski space. See Fig.~\ref{Fig1}-(b) for the shapes of the warp factor. 

Note that asymmetric gravitating kink solutions were first discussed in some 5D thick brane models~\cite{EtoSakai2003,TakamizuMaeda2006,BazeiaMenezesRocha2014}. Such solutions are especially important for simulating the collision of gravitating kinks, see Ref.~\cite{TakamizuMaeda2006}.

\section{Linear perturbation}
The linear perturbation issue for arbitrary solutions of action~\eqref{1} with metric \eqref{metricXCord} has been discussed in Refs.~\cite{Zhong2021,Zhong2021b}. Here we only briefly review the key steps and then apply the results to the above double-kink solution.

The first step is to introduce a new spatial coordinate
\begin{equation}
r\equiv\int e^{-A(x)}dx,
\end{equation}
with which the line element becomes conformally flat
\begin{equation}
  ds^2=e^{2A(r)}\big(-dt^2+dr^2\big).
\end{equation}
For simplicity, we will use overdots and primes to denote the derivatives with respect to $t$ and $r$, respectively.

Then, we consider small perturbations around the background solutions
\begin{eqnarray}
g_{\mu\nu}(r)&+&\delta  g_{\mu\nu}(t, r),\\
 \varphi(r)&+&\delta \varphi(t,r),\\
\phi(r)&+&\delta\phi(t,r),
\end{eqnarray}
where the metric perturbation is defined as the following form:
\begin{eqnarray}
\delta g_{\mu\nu}(r,t)&\equiv& e^{2A(r)} h_{\mu\nu}(r,t)\nonumber\\
&=&e^{2A(r)} \left(
\begin{array}{cc}
 h_{00}(r,t) & \Phi (r,t) \\
 \Phi (r,t) & h_{rr}(r,t) \\
\end{array}
\right).
\end{eqnarray}

As  in Ref.~\cite{Zhong2021}, we define a new variable $\Xi \equiv 2 \dot{\Phi}-h_{00}^{\prime}$, and conduct our calculations in the dilaton gauge, where $\delta \varphi=0$. It is worth to mention that the dilaton gauge is an appropriate gauge, as it leads to the correct equation for the normal modes of the physical perturbation~\cite{Zhong2021b}. 

After linearizing the field equations and using the dilaton gauge, one obtains three independent linear perturbation equations. Two of them come from the $(0,1)$ and the $(1,1)$ components of the linearized Einstein equations, which are
\begin{eqnarray}
\label{eqPer1}
\varphi '{h_{rr}}=\kappa \phi '\delta \phi,
\end{eqnarray}
and
\begin{eqnarray}
\label{eqPer2}
&&-2A'\varphi 'h_{rr}
+\frac{1}{2}h_{rr}\varphi '^2+\Xi \varphi '\nonumber\\
&=&\kappa \left( \phi '\delta \phi '-\phi ''\delta \phi -\frac{1}{2}\phi '^2h_{rr} \right),
\end{eqnarray}
respectively. The third one comes from the linearized scalar field equation
 \begin{eqnarray}
\label{PertEq3}
&&-\ddot{\delta}\phi +\delta \phi ''+2A'\frac{\phi ''}{\phi '}\delta \phi -\frac{\phi '''}{\phi '}\delta \phi \nonumber\\
&&-\frac{1}{2}\phi ' h_{rr}'-\phi ''h_{rr}+\frac{1}{2} \phi '\Xi=0,
\end{eqnarray}
which, after eliminating $h_{rr}$ and $\Xi$ by using Eqs.~\eqref{eqPer1} and \eqref{eqPer2}, becomes a wave equation of $\delta\phi$~\cite{Zhong2021}:
\begin{eqnarray}
\label{eqDelPhi}
\ddot{\delta \phi }-\delta \phi ''+  V_{\text{eff}}(r)\delta \phi=0.
\end{eqnarray}
Here, the effective potential is defined by
\begin{eqnarray}
V_{\text{eff}}(r)\equiv \frac{f ''}{f}, \quad \text{with} \quad f\equiv \frac{\phi '}{\varphi '}.
\end{eqnarray}
If we take $\delta\phi=\psi(r)e^{iwt}$, Eq.~\eqref{eqDelPhi} becomes a Schr\"odinger-like equation of $\psi(r)$:
\begin{eqnarray}
-\psi ''+  V_{\text{eff}}  \psi=w^2 \psi.
\end{eqnarray}
An interesting observation is that the Hamiltonian operator is factorizable:
\begin{eqnarray}
\hat{H}=-\frac{d^2}{dr^2}+V_{\text{eff}}=\hat{\mathcal{A}}\hat{\mathcal{A}}^\dagger,
\end{eqnarray}
where
\begin{eqnarray}
\mathcal{A}=\frac{d}{d r}+\frac{{f}'}{f}, 
~ \mathcal{A}^{\dagger}=-\frac{d}{d r}+\frac{{f}'}{f}.
\end{eqnarray}

If $V_{\text{eff}}(r)$ is a smooth function, one can state that the eigenvalues of a factorizable Hamiltonian operator are semipositive definite, namely, $w^2\geq 0$, see~\cite{InfeldHull1951,CooperKhareSukhatme1995} for details. In this case, we say that the static background solution is stable against linear perturbations. The zero mode ($w_0=0$) satisfies $ \mathcal{A}^{\dagger}\psi_0(r)=0$, and the solution reads
\begin{eqnarray}
\psi_0(r)\propto f=\frac{\phi '}{\varphi '}=\frac{\phi '}{2A '}.
\end{eqnarray}

\begin{figure}
\centerline{\includegraphics[width=8cm]{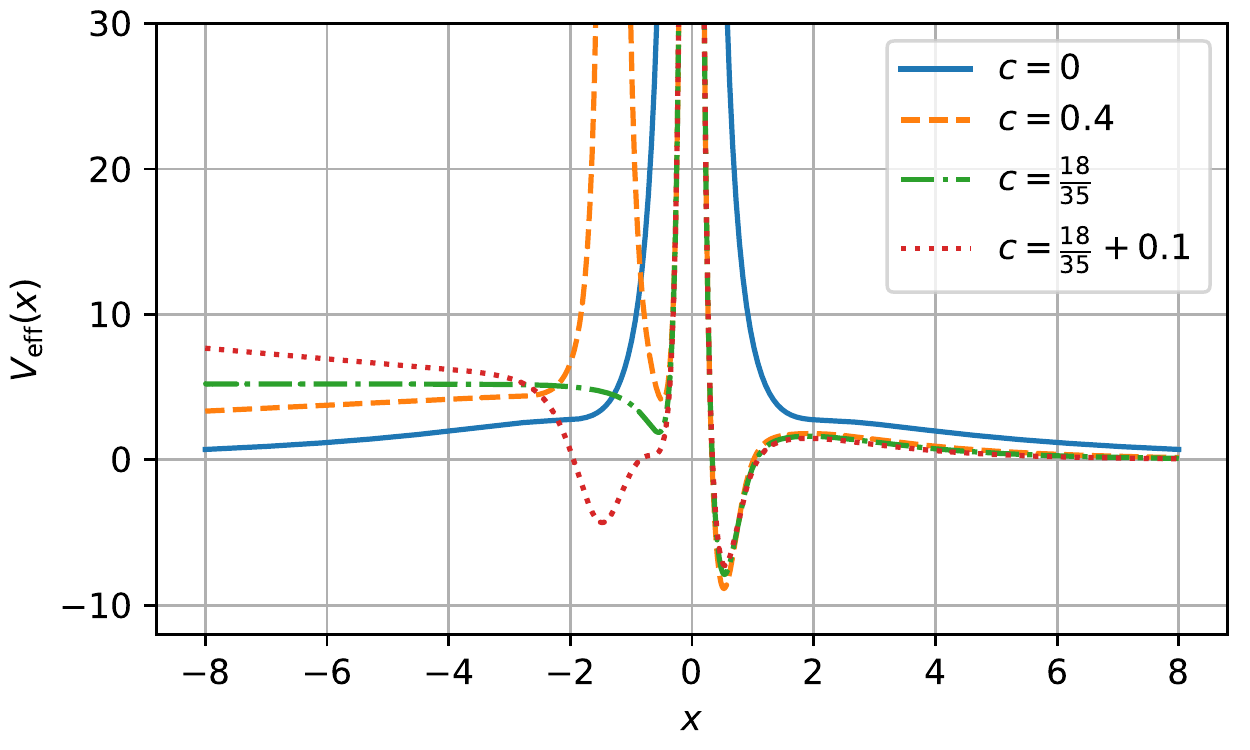}}
\caption{The effective potential for $p=3$, $\kappa=1$. For both $c=0$ and $c>c_{\textrm{cr}}$ there is only one repulsive singularity locates at $x=0$. While when $0<c\leq c_{\textrm{cr}}$ there is another singularity at $x=x_s$. For $c=0.4$, $x_s\approx-1.29505$.}
\label{FigVeff}
\end{figure}

In general, it is very hard to obtain the analytical expressions of $V_{\text{eff}}(r)$ and $\psi_0(r)$, because this requires an explicit expression of $x(r)$, which can only be obtained if the warp factor takes some special forms. Thus, it will be useful  to transform $V_{\text{eff}}$  back to the $x$-coordinates:
\begin{eqnarray}
V_{\text{eff}}(x)&=&e^{2A}\left(\partial_x A\frac{\partial_x f}{f}+\frac{\partial_x^2f}{f}\right),
\end{eqnarray}
with $f(x)=\frac{\partial_x\phi}{2\partial_x A}$. After substituting the double-kink solution~\eqref{solKink} and \eqref{SolWarp} into the above equation, we immediately obtain a lengthy but analytical expression of $V_{\text{eff}}(x)$, whose shapes can be found in Fig.~\ref{FigVeff}.
\begin{figure}
\centerline{\includegraphics[width=8cm]{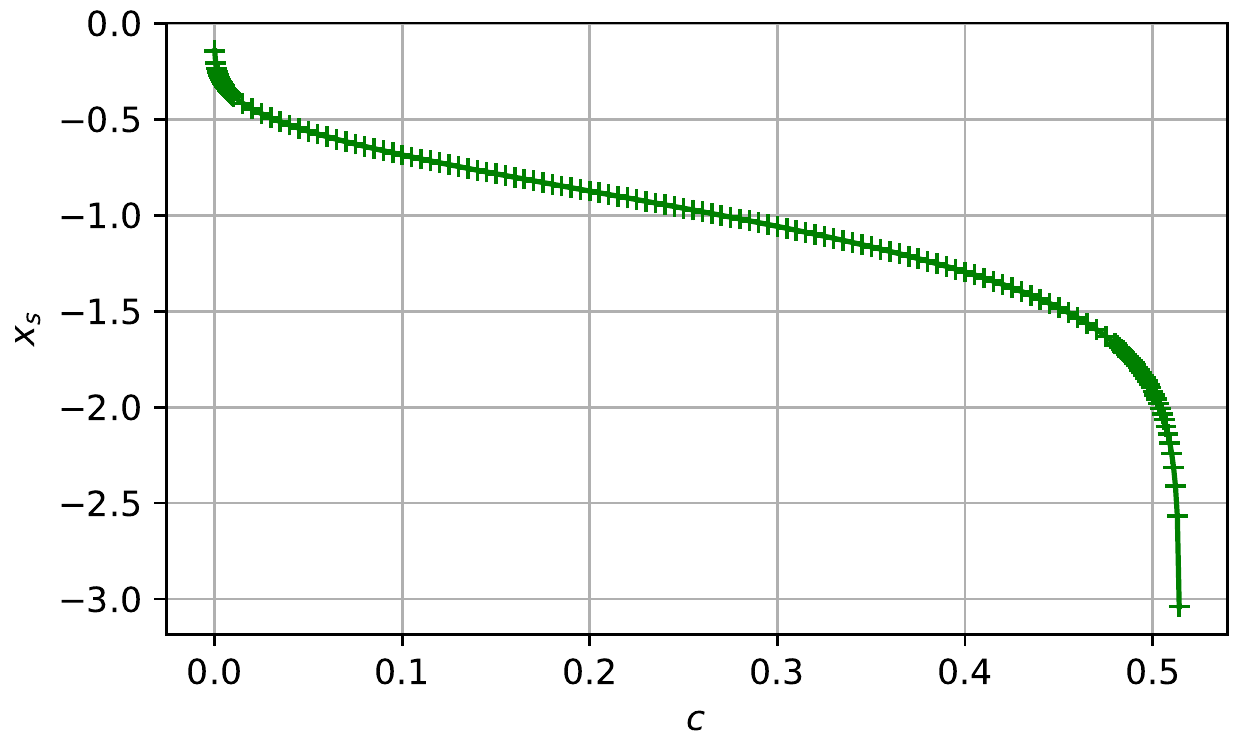}}
\caption{Numerical results of the location of the singularity at $x_s$ as a function of paramter $c$. Clearly, $x_s\to 0$ as $c\to 0$, and  $x_s\to -\infty$ as $c\to c_{\textrm{cr}}\approx 0.5143$.}
\label{Figxs}
\end{figure}
We see that when $c=0$ there is only one repulsive singularity at $x=0$. But for $0<c\leq c_{\textrm{cr}}$, there are two repulsive singularities. One singularity locates at $x=0$, the other at some $x_s<0$. One can show that $x_s$ corresponds to the extreme of the warp factor, and satisfies the following equation:
\begin{equation}
\label{eqxs}
35 c+9 \tanh ^5(x_s) \left(5 \text{sech}^2(x_s)+2\right)=0.
\end{equation}
The root of this equation can be found out numerically. Our calculation shows that $x_s\to 0$ as $c\to 0$, and  $x_s\to -\infty$ as $c\to c_{\textrm{cr}}$, see Fig.~\ref{Figxs}.  

As can be see from Fig.~\ref{FigVeff}, for $c\in[0, c_{\textrm{cr}}]$ the effective potential $V_{\text{eff}}$ is always positive everywhere, and all the eigenvalues are non-negative. Therefore, the double-kink solution is stable against linear perturbation in this case.

On the other hand, if $c> c_{\textrm{cr}}$, Eq.~\eqref{eqxs} has no real root, and the number of the singularities reduces to one again. But in this case, some potential well appears around $x=0$, where $V_{\text{eff}}$ can be negative.  Interestingly, however, the zero mode wave function
\begin{equation}
\label{eqPsi0}
\psi_0(x)\propto \frac{210 \tanh ^2(x) \text{sech}^2(x)}{35 c \kappa +9 \kappa  \tanh ^5(x) \left(5 \text{sech}^2(x)+2\right)}
\end{equation}
becomes normalizable
 \begin{equation}
\mathcal{N}^2\int_{-\infty}^{+\infty}dx e^{-A}\psi_0^2(x)=1,\quad \mathcal{N}<\infty.
\end{equation}
For example, for $\kappa=1$, $c=0.6, 0.8$ and $1$ numerical calculations show that $\mathcal{N}\approx 0.162, 0.320$ and 0.445.
\begin{figure}
\centerline{\includegraphics[width=8cm]{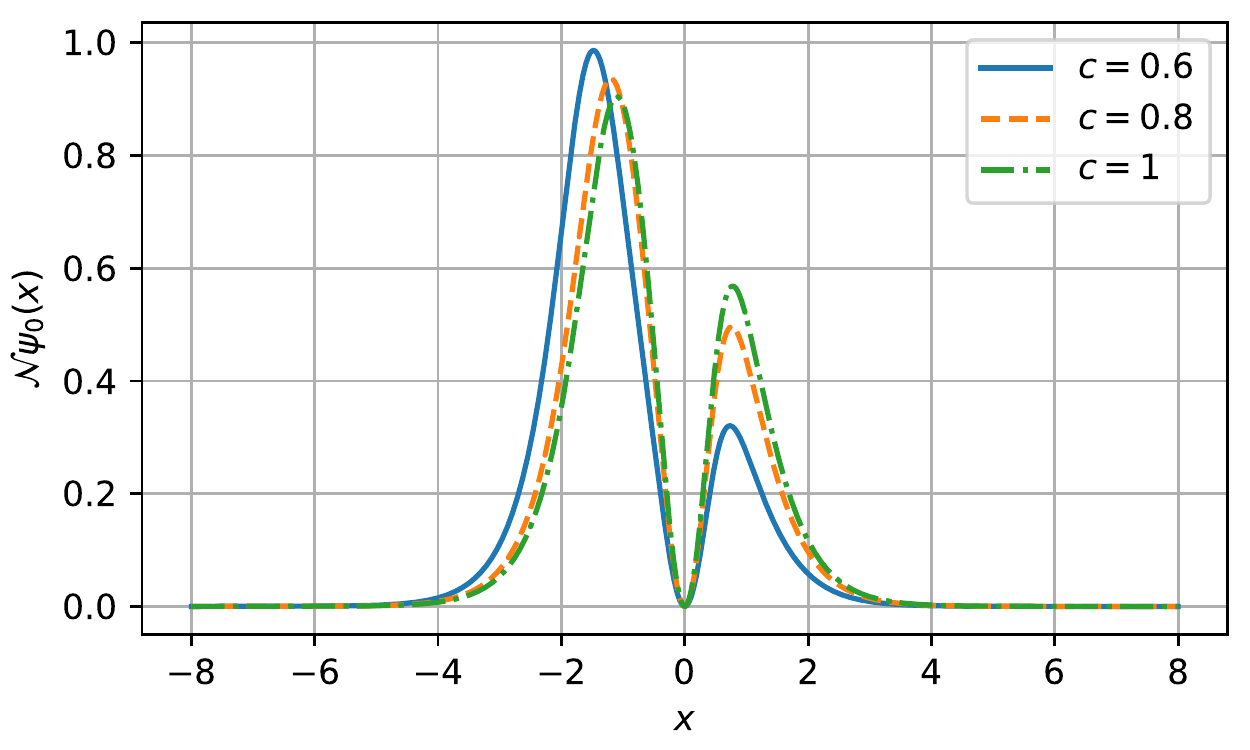}}
\caption{The normalized zero mode wave functions for $p=3$, $\kappa=1$, $c=0.6, 0.8$ and $1$.}
\label{FigPsi0}
\end{figure}
From Fig.~\ref{FigPsi0} we see that the wave functions of the zero mode are vanished at $x=0$, due to the presence of the repulsive singularity. The absent of any node in $\psi_0(x)$ indicates that the zero mode is the ground state of the Hamiltonian operator. Therefore, the double-kink solution is also stable for $c> c_{\textrm{cr}}$.

\section{Conclusion and discussions}
In this work, we studied the linear perturbation issue of static gravitating double-kink solutions. For simplicity, we constructed a 2D model with analytical double-kink solutions, and discussed the behavior of the effective potential of the linear stability equation. The result should reflect some properties of the scalar perturbation of the corresponding 5D Einstein-scalar model~\cite{BazeiaFurtadoGomes2004}.

Different from Ref.~\cite{BazeiaFurtadoGomes2004}, in the present work we introduced a parameter $c\geq0$, which allows the metric to be both symmetric ($c=0$) and asymmetric ($c>0$). We found that double-kinks with asymmetric warp factors can have two repulsive singularities in its effective potential if $0<c\leq c_{\textrm{cr}}=18/35$. One of the singularities locates at $x=0$, while the other at some $x_s(c)<0$, whose value can be determined numerically. Especially, when $c\to 0$, $x_s\to 0$, while when $c\to c_{\textrm{cr}}$, $x_s\to-\infty$. This is a new feature which has not been observed in single kink  models~\cite{Zhong2021,ZhongLiLiu2021}, where there is at most one repulsive singularity.

For both $c=0$ and $c>c_{\textrm{cr}}$ there is only one repulsive singularity locates at $x=0$. Despite of the presence of the singularity, the zero mode wave function $\psi_0(x)$ is normalizable for the later case. The fact that $\psi_0(x)$ is vanished at $x=0$ indicates that the singularity is a strong one~\cite{CooperKhareSukhatme1995}. A strong singularity at $x=0$ forces $\psi(0)=0$ and the range of $x$ is effectively broken into two disjoint pieces $x<0$ and $x>0$ with no communication between them~\cite{CooperKhareSukhatme1995}. Note that for the single kink case~\cite{Zhong2021,ZhongLiLiu2021}, if the asymmetric parameter is large enough, the zero mode also becomes normalizable. But in that case $\psi(0)\neq0$, as no singularity appears at $x=0$.

We also stated that the present double-kink solution is always stable against linear perturbations. Because when $0<c\leq c_{\textrm{cr}}$ the effective potential $V_{\textrm{eff}}$ is positive everywhere, while when  $c> c_{\textrm{cr}}$, the normalizable zero mode is the ground state.

The present work shows that gravitating double-kink solutions might have very different scalar perturbation spectra than single kinks, and deserve further investigations. 

Besides, the present work as well as Refs.~\cite{Zhong2021,ZhongLiLiu2021} all indicate that singular potentials are closely related to the scalar perturbation issue of gravitating kink and double-kink solutions.  Unfortunately, this simple fact did not receive enough attentions. This is because in many gravitating kink models, the scalar perturbation is discussed by directly linearizing the field equations under the longitude gauge, for example, see Ref.~\cite{KobayashiKoyamaSoda2002}. With the longitude gauge, one can also obtain a simple Schr\"odinger-like equation with Hamiltonian operator, say $\hat{H}_-$. But as have been stated in Refs~\cite{Giovannini2003,Giovannini2001a,ZhongLiu2013,Zhong2021b}, the Hamiltonian operator for the normal modes of the physical perturbation, say $\hat{H}$, should be derived from the quadratic action of the perturbations.
Interestingly, $\hat{H}_-$ and $\hat{H}$ are superpartners in the sense of supersymmetric quantum mechanics, and their spectra are related, provided that their potentials are non-singular. However, if one of the Hamiltonian operators has a singular potential, as in the gravitating kink/double-kink cases, the spectra of $\hat{H}_-$ and $\hat{H}$ are different in general~\cite{JevickiRodrigues1984,CasahorranNam1991,Casahorran1991,PanigrahiSukhatme1993,DasPernice1999}. Consequently, conclusions based on $\hat{H}_-$ might be incorrect.

The present work as well as Refs.~\cite{Zhong2021,ZhongLiLiu2021,Zhong2021b} all assume that the metrics are static. In some interesting 5D Einstein-scalar models, it is also possible to construct gravitating kink solutions where the metrics depend on both space and time coordinates~\cite{LiuZhongYang2010,GermanHerreraAguilarMalagonMorejonMoraLunaEtAl2013,GermanHerreraAguilarKuertenMalagonMorejonEtAl2016,BarbosaCendejasCartasFuentevillaHerreraAguilarMoraLunaEtAl2018a,BarbosaCendejasCartasFuentevillaHerreraAguilarMoraLunaEtAl2018}. These solutions describe thick de-Sitter brane worlds. It is natural to ask if the corresponding 2D models can be constructed? We leave this interesting question to our future works.

\acknowledgments
 This work was supported by the National Natural Science Foundation of China (Grant Nos.: 12175169, 12075178), Fundamental Research Funds for the Central Universities (Grant Number: xzy012019052) and Natural Science Basic Research Plan in Shaanxi Province of China (No. 2018JM1049).

%\bibliographystyle{eplbib}
%\bibliography{/Users/zhongyuan/360Yun/jabref/library/articles}

\end{document}